\documentclass[preprint,5p,times,sort&compress,twocolumn]{elsarticle}
\usepackage{amsthm, amsfonts, amssymb, comment}	%%
\usepackage[fleqn]{amsmath}				%%
\usepackage{graphicx}					%%
\usepackage{txfonts}					%%
\usepackage{hyperref,xcolor}			%%
\usepackage{geometry}                   %%
\usepackage{lineno}						%%
\modulolinenumbers[1]					%%
\usepackage{mathtools, cuted}           %%
\newcommand{\newc}{\newcommand}
\newc{\beq}{\begin{equation}}
\newc{\eeq}{\end{equation}}
\newc{\bea}{\begin{array}}
	\newc{\eea}{\end{array}}
\newcommand{\ben}{\begin{eqnarray}}
\newcommand{\een}{\end{eqnarray}}
\newc{\ra}{\rightarrow}
\newc{\bfx}{{\bf x}}
\newc{\bfV}{{\bf V}}
\newc{\cO}{{\cal O}}
\newc{\bfv}{{\bf v}}
\newc{\bfu}{{\bf u}}
\newc{\bfp}{{\bf p}}
\newc{\ve}{{\varepsilon}}
\newc{\Psibar}{\overline\Psi}
\newc{\w}{{\bf w}}
\newc{\E}{{\mathbf{E}}}
\newc{\EE}{{\mathcal E}}
\newc{\bfn}{{\mathbf\nabla}}
\newc{\la}{{\cal L}}
\newc{\tla}{{\tilde{\cal L}}}
\newc{\bp}{{\bf p}}
\newc{\ho}{\hookrightarrow }
\newc{\bP}{{\bf P}}
\newc{\pd}{{\partial}}
\newc{\piv}{{\partial_4}}
\newc{\pv}{{\partial_5}}
\newc{\bJ}{{\bf J}}
\newc{\bze}{{\mathbf 0}}
\newc{\bK}{{\bf K}}
\newc{\tphi}{{\tilde\phi}}
\newc{\tF}{{\tilde F}}
\newc{\tD}{{\tilde D}}
\newc{\tJ}{{\tilde J}}
\newc{\tj}{{\tilde j}}
\newc{\bD}{{\bf D}}
\newc{\tvphi}{{\tilde\varphi}}
\newc{\trho}{{\tilde\rho}}
\newc{\ttheta}{{\tilde\theta}}
\newc{\tpsi}{{\tilde\psi}}
\newc{\tu}{{\tilde u}}
\newc{\cD}{{\cal D}}
\newc{\tPhi}{{\tilde\Phi}}
\newc{\tPsi}{{\tilde\Psi}}
\newc{\tA}{{\tilde A}}
\newc{\talpha}{{\tilde\alpha}}
\newc{\tbeta}{{\tilde\beta}}
\newc{\bA}{{\mathbf A}}
\newc{\bB}{{\bf B}}
\newc{\br}{{\bf r}}
\newc{\sig}{{\mathbf\sigma}}
\newc{\eg}{{\rm e.g.\ }}
\newc{\ie}{{\rm i.e.\ }}
\newcommand{\bey}{\begin{eqnarray}}
\newcommand{\pslash}{\not{\hbox{\kern-2.3pt $p$}}}
\newcommand{\pdslash}{\not{\hbox{\kern-2pt $\partial$}}}
\newcommand{\eey}{\end{eqnarray}}

\begin{document}
	
	\begin{frontmatter}
		\title{Morse potential in relativistic contexts from generalized momentum operator, Pekeris approximation revisited
        and mapping
%from generalized momentum operator and
%non-minimal coupling
}
		
		\author[salvador]{Ignacio S. Gomez\corref{cor1}}
		\ead{nachosky@fisica.unlp.edu.ar}
		\author[salvador]{Esdras S. Santos}
		\ead{esdras.santos@ufba.br}
		\author[salvador]{Olavo Abla}
		\ead{olavo.abla@ufba.br}
		\cortext[salvador]{Corresponding author}
		\address[salvador]{Instituto de F\'{i}sica, Universidade Federal da Bahia,
			Rua Barao de Jeremoabo, 40170-115 Salvador-BA, Brazil}

\begin{abstract}
In this work we explore a generalization of the Dirac and Klein-Gordon (KG) oscillators,
provided with a deformed linear momentum
%is replaced by a deformed version
inspired in nonextensive statistics, that gives place to the Morse potential
in relativistic contexts by first principles.
%We employ a coordinate transformation that maps
In the (1+1)--dimensional case the relativistic oscillators
%with generalized momentum operator
are mapped into the quantum Morse potential.
%We analyze the one-dimensional and three-dimensional cases, using the Pekeris approximation in the later one.
Using the Pekeris approximation, in the (3+1)--dimensional case we study
the thermodynamics of the S-waves states ($l=0$) of the H$_2$, LiH, HCl and CO molecules
(in the non-relativistic limit) and of a relativistic electron,
%in the high energy relativistic regime,
where Schottky anomalies 
(due to the finiteness of the Morse spectrum) and
spin contributions to the heat capacity are reported.
By revisiting a generalized Pekeris approximation, we provide a mapping from (3+1)--dimensional Dirac and KG equations
with a spherical potential to an associated one-dimensional Schr\"odinger-like equation, and we obtain
the family of potentials for which this mapping corresponds to a Schr\"odinger equation with non-minimal coupling.
\end{abstract}

		
	\end{frontmatter}
	
	\nolinenumbers   %package lineno
	
%%%%%%%%%%%%%%%%%%%%%%%%%%%%%%%%%%%%%%%%%%%%%%%%%%%%%%%%%%%%%%%%%%%%%%
\section{Introduction}
%%%%%%%%%%%%%%%%%%%%%%%%%%%%%%%%%%%%%%%%%%%%%%%%%%%%%%%%%%%%%%%%%%%%%%

For compatibilizing the principles of quantum mechanics with the special relativity,
the Klein-Gordon (KG) and the Dirac equations constitute the two most relevant
cases that
were found, both satisfying the quadratic relativistic relation $E^2=p^2c^2+m^2c^4$.
The KG equation is obtained by means of the
quantization this relation, which leads to a relativistic
wave equation with second order derivatives in time and space that is
Lorentz-covariant, while Dirac focused on a relativistic first order wave equation
describing the behavior of electrons consistently
%in the atom in a consistent manner
with the special relativity \cite{Dirac}.

In Refs. \cite{Moshinsky-1989,Bruce,Comment}
an harmonic potential has been incorporated by
adding to the linear momentum (non-minimum coupling) a linear function,
thus obtaining the so called Dirac and KG oscillators,
that in the non-relativistic limit
gives the quantum harmonic oscillator for spinless and
strong spin-orbit coupling fermionic particles.
These types of linear interactions
were employed in quarks mass spectra \cite{Kang-1975}, on a Coulomb-like
potential \cite{Diracoulomb,KGCoulomb}, in $2D$ massless fermions \cite{2D}
and propagators \cite{Propagator}, in curved space-time \cite{Topology},
in systems with extended and generalized uncertainty principle \cite{EUP,GUP}.
Also, the non-relativistic quantum-mechanical formalism was studied by
some authors \cite{Costa_Filho-2011,Costa_Filho-2013,EUP2016}, that have considered the harmonic
oscillator provided with a generalized
linear momentum operator which allows to
obtain the Morse potential \cite{Morse-1929} by first principles.
Recently, some of us have characterized
a deformed lattice using the same generalized
linear momentum operator \cite{Costa-Gomez-2020}.

The goal of this letter is twofold. First, we extend the strategy
used in \cite{Costa_Filho-2013}
to the one-dimensional and three-dimensional KG and Dirac oscillators
in order to obtain
the corresponding relativistic equations for
a Morse potential coupling \cite{Alhaidari,CommentDIRACMORSE,Alhaidari2,
Chen,KGDMORSE,Berkdemir,Qiang-2007,Zhang-2009,
KGMORSESOLUTION,Oyewumi,Ortakaya,KGMORSE,Zhang-2016,Garcia,Castro,Last}.
Then, from a generalized Pekeris approximation we obtain a mapping
between (3+1)-dimensional KG and Dirac equations and Schr\"odinger-like ones for
arbitrary spherical potentials.
The work begins with the
%theoretical
preliminaries, with the development of a
generalized linear momentum operator formulation
in the Hamiltonian, showing that both oscillators
with this coupling are equivalent the one-dimensional KG and Dirac
particles provided with a Morse potential coupling
and the standard linear momentum. We illustrate the results
with those obtained in the literature for the $\textrm{H}_2$ molecule \cite{Nasser-2007,H2,DISSOCIATION}.
Using the Pekeris approximation \cite{Pekeris-1933} in the three-dimensional case
for the Morse potential non-minimal coupling,
we obtain the eigenvalues and the
eigenfunctions,
%the dissociation energy
%function of the quantum numbers,
and then we recover
the non-relativistic and the non-deformed limits for both cases.
Hence, in order to test our approximations in the three-dimensional case, we
study the thermodynamics of the S-wave states ($l=0$) for the
H$_2$, LiH, HCL and CO molecules (in the non-relativistic limit)
and of an electron in the high energy relativistic regime.
Here, Schottky anomalies are reported in the heat capacity as a consequence of
the spectrum finiteness of the Morse potential mapping.
Next, we revisit the generalized Pekeris approximation \cite{Pekeris2}
to provide
a mapping for passing from (3+1)-dimensional KG and Dirac equations
with an arbitrary spherical
non-minimal coupling to an associated Schr\"odinger-like equation. Also, we
determine the family of potential couplings from which
the mapping is performed onto a Schr\"odinger equation with non-minimal coupling.
Finally, we outline our conclusions.

\section{Preliminaries}

We present the preliminaries used throughout the work.

\subsection{Morse potential and generalized momentum operator}\hspace{0cm}
Some authors
\cite{Costa_Filho-2011,Costa_Filho-2013,EUP2016,
Mazharimousavi-2012,Costa-Borges-2014,Costa-Gomez-2018,Costa-Gomez-2020}
have investigated a generalized translation operator
that gives a nonadditive spatial displacement of the form
\begin{equation}
\label{eq:T_gamma}
	\hat{\mathcal{T}}_{\gamma}(\varepsilon)|x \rangle
			= | x + \varepsilon + \gamma x \varepsilon \rangle
\end{equation}
being $\varepsilon$ an infinitesimal displacement and $\gamma$
a parameter with dimension of inverse length in such a way that
$\gamma_q \equiv (1-q)/\xi$ (from now on we place
implicitly the dependence on $q$ in $\gamma$)
with $\xi$ the characteristic
length of the system, where the usual translation
is recovered for $q \rightarrow 1$ ($\gamma \rightarrow 0$).
These investigations were inspired by the
development of Tsallis nonextensive
thermostatistics \cite{Tsallis-1988,Tsallis-1999,TsallisPRL} along with some of
its mathematical implications (the $q$-calculus \cite{Ernesto}).
In Ref. \cite{Costa_Filho-2011,Costa_Filho-2013,EUP2016}
by means of $\hat{\mathcal{T}}_{\gamma}(\varepsilon)$
the authors obtained the generalized momentum operator
$\hat{p}_\gamma |\alpha\rangle=-i\hbar D_\gamma|\alpha\rangle$
in the basis $x$,
being $D_\gamma$ the deformed derivative in $x$
\begin{equation}\label{deformed-derivative}
D_\gamma=(1+\gamma x)\frac{d}{dx}.
\end{equation}
By means of the Hamiltonian
$H=\hat{p}_\gamma^2+V(x)$ and using \eqref{deformed-derivative}
it follows the
(deformed) Schr\"odinger-like equation
%\eqref{deformed-SE}
\begin{equation}\label{deformed-SE}
i\hbar \frac{\partial}{\partial t}\psi(x,t)=
-
%\left(
\frac{\hbar^2}{2m}
%\right)
D_\gamma^2\psi(x,t)+V(x)\psi(x,t)
\end{equation}
that corresponds to a particle provided
with an effective mass $m(x)=m/(1+\gamma x)^2$.
Eq. \eqref{deformed-SE} has been employed in applications of
semiconductor heterostructures
\cite{BenDaniel-Duke-1966,Serra-Lipparini-1997}.

An interesting application of the deformed Schr\"odinger equation
\eqref{deformed-SE} was given in order to derive the Morse potential
by using first principles \cite{Costa_Filho-2013}. More precisely,
by considering the coordinate transformation
\begin{equation}\label{canonical}
\eta=\frac{\ln(1+\gamma x)}{\gamma}
\end{equation}
in \eqref{deformed-SE} along with the
harmonic potential $V(x)=m\omega^2x^2/2$  and
$\phi(\eta,t)=\psi(x(\eta),t)$
the following equation is obtained (using $E=i\hbar \frac{\partial}{\partial t}$)
\begin{equation}\label{Morse-SE}
E\phi(\eta,t)=
-\frac{\hbar^2}{2m}\frac{d^2}{d\eta^2}\phi(\eta,t)+
\frac{m\omega^2}{2\gamma^2}(e^{\gamma \eta}-1)^2\phi(\eta,t),
\end{equation}
which is precisely the
eigenvalues equation of the quantum
Morse oscillator (QMO) \cite{Morse-1929}. This is provided
of an effective potential
$V_{\textrm{eff}}(\eta)=D(e^{\gamma \eta}-1)^2$,
for the wave function $\phi(\eta,t)$ in
the $\eta$ space with
the dissociation parameter $D=\frac{m\omega^2}{2\gamma^2}$.
The eigenfunctions of the QMO are given by
%\cite{Morse-1929}
\begin{equation}\label{eigenfunctions-QMO}
\Phi_n(z)=A_n z^se^{-\frac{1}{2}z}L_n^{2s}(z)
\end{equation}
with $A_n$ the normalization constant,
$z=2m\omega e^{\gamma \eta}/(\gamma^2\hbar)$,
$s=m\omega/(\gamma^2\hbar)-n-1/2$ and
$L_n^{2s}(z)=(z^{-2s}e^z/n!)d^n(e^{-z}z^{n+2s})/dz^n$
the generalized Laguerre polynomial \cite{Arfken-2005}.
The energy spectrum of the QMO is
\begin{equation}\label{eigenergies-QMO}
E_n=\hbar\omega\left(n+\frac{1}{2}\right)
\left[1-\frac{\gamma^2\hbar}{2m\omega}\left(n+\frac{1}{2}\right)\right]
\end{equation}
being $n$ restricted to the range $0\leq 2n\leq 2m\omega/(\gamma^2\hbar)-1$,
which implies a finite number of states and $E_n\geq \hbar\omega(n+1/2)/2$.
From \eqref{eigenergies-QMO} it can be seen that
the harmonic oscillator energies
are recovered for $\gamma\rightarrow0$.

\subsection{Klein-Gordon and Dirac oscillators}\hspace{0cm}
The substitution of the four-vector energy-momentum
$\mathbf{p}^\mu=(E/c,\textbf{p})=(i\hbar\partial/\partial t,-i\hbar\nabla)$
in the quadratic relativistic relation
for the case of an harmonic coupling prescription
gives the
\emph{Klein-Gordon oscillator} \cite{Bruce,Comment}
\begin{equation}\label{KG-oscillator}
2m\mathcal{E}\psi=(\textbf{p}+im\omega\textbf{r})\cdot(\textbf{p}-
im\omega\textbf{r})\psi,
\end{equation}
where the form $(\textbf{p}+im\omega\textbf{r})\cdot(\textbf{p}-
im\omega\textbf{r})$ ensures the Hermiticity \cite{Boumali-2011}.
From now on, for practical and notation reasons
we will denote
$\mathcal{E}$ as $\frac{E^2-m^2c^4}{2mc^2}$ with $E$ the energy of the particle.
Here $\omega$ is the frequency of the oscillator, $m$ the
mass and $\textbf{r}$ the position,
with the limit $\omega\rightarrow0$ the corresponding one to
the free particle case.
Using algebraic methods the energies $E_N$
for the one-dimensional case result \cite{Boumali-2011}
\begin{equation}\label{KG-energies}
\mathcal{E}_N=
N\hbar\omega
\quad , \quad N=0,1,2,\ldots
\end{equation}
with $\mathcal{E}_N=(E_N^2-m^2c^4)/2mc^2$. In the non-relativistic limit
$E=mc^2+\epsilon$ with $\epsilon\ll mc^2$ and then we have
$\epsilon_N\approx N\hbar\omega$ ($N=0,1,2,\ldots$)
that correspond to the energies of the harmonic oscillator provided with a zero ground state energy.

The usual form of the Dirac equation for a particle of mass $m$ is given by
\begin{eqnarray}\label{Dirac}
\left[i\hbar\beta\frac{\partial}{\partial t}+i\hbar\beta\overrightarrow{\alpha}\cdot\overrightarrow{\nabla}-\frac{mc}{\hbar}\right]\psi=0
\end{eqnarray}
where $\beta$ and $\overrightarrow{\alpha}$ provide the $(3+1)$ representation of the Dirac matrices
\begin{equation}
\beta=
\begin{pmatrix}
\mathbf{I} & 0 \\
0 & \hspace{-0.3cm}-\mathbf{I}
\end{pmatrix};\hspace{0.5cm}\alpha^i=
\begin{pmatrix}
0 & \sigma^i \\
\sigma^i & 0
\end{pmatrix};\nonumber
\end{equation}
and $\mathbf{I}$ is the $2\times2$ identity with $\sigma^1,\sigma^2$ and $\sigma^3$ the $2\times2$ Pauli matrices.
Using the non-minimal harmonic coupling $\textbf{p}-im\beta\omega\textbf{r}$
in \eqref{Dirac} we obtain the \emph{Dirac oscillator} \cite{Moshinsky-1989}
\begin{eqnarray}\label{Dirac-oscillator}
&(E-mc^2)\psi_1=
c\overrightarrow{\sigma}\cdot(\textbf{p}+im\omega \textbf{r})\psi_2 \nonumber \\
&(E+mc^2)\psi_2=
c\overrightarrow{\sigma}\cdot(\textbf{p}-im\omega \textbf{r})\psi_1
\end{eqnarray}
where $\omega$ is the frequency of the
oscillator and
$\psi^T=(\psi_1,\psi_2)$  is the spinorial wavefunction.
From \eqref{Dirac-oscillator} it follows the differential equation for
$\psi_1$
\begin{eqnarray}\label{Dirac-oscillator-DE}
\mathcal{E}\psi_1=\left[\frac{p^2}{2m}+\frac{m\omega^2r^2}{2}-\frac{3}{2}\hbar\omega-
\frac{2\omega}{\hbar}
\mathbf{L}\cdot\mathbf{S}\right]\psi_1
\end{eqnarray}
where $\mathbf{L}=\mathbf{r}\times\mathbf{p}$
is the angular momentum and
$\mathbf{S}=(\hbar/2)\sigma$ is the spin operator.
By means of the total spin  $\mathbf{J}=\mathbf{L}+\mathbf{S}$
it can be shown that the energies $E_{Nlj}$ are \cite{Moshinsky-1989}
\begin{eqnarray}\label{Dirac-energies}
&E_{Nlj}^2-m^2c^4=\hbar\omega[2(N+1-j)\mp1]mc^2,\\
&\hspace{4cm}\quad \textrm{if} \quad l=j\mp\frac{1}{2}\nonumber
\end{eqnarray}
which presents a degeneracy (typically of central potentials) for the pairs
$(N\pm1,j\mp1),(N\pm2,j\mp2),\ldots$.
From \eqref{Dirac-energies} in
the non-relativistic limit
$E=mc^2+\epsilon$ with $\epsilon\ll mc^2$ we have
$\mathcal{E}=\epsilon_{Nlj}\approx \hbar\omega(N+1-j\mp\frac{1}{2})$ for $N=0,1,2,\ldots$
and $l=j\mp\frac{1}{2}$, so we recover the energies of the harmonic oscillator energies
plus a strong spin-orbit term.

\section{Morse potential for Klein-Gordon and Dirac equations from generalized momentum couplings}

We present the Klein-Gordon and Dirac equations with Morse potential from the
Klein-Gordon and Dirac oscillators provided with a generalized momentum coupling.
We consider the one-dimensional and the three-dimensional cases.

\subsection{One-dimensional case}\hspace{0cm}
Considering the Eq. \eqref{KG-oscillator} in one dimension
with the deformed derivative \eqref{deformed-derivative} we have
\begin{eqnarray}
\label{KG-Morse-oscillator-1D-1}
2m\mathcal{E}\psi(x)=(p_{\gamma}+im\omega x)(p_{\gamma}-im\omega x)\psi(x),
\end{eqnarray}
that can be considered the Klein-Gordon version
of the generalized harmonic oscillator studied in \cite{Costa_Filho-2011,Costa_Filho-2013}.
From \eqref{KG-Morse-oscillator-1D-1}
and using the coordinate transformation \eqref{canonical} we have
\begin{eqnarray}
\label{KG-Morse-oscillator-1D-2}
\mathcal{E}\phi(\eta)=\frac{1}{2m}\Bigg{\{}\left[-i\hbar\frac{d}{d\eta}+
im\omega\left(\frac{e^{\gamma \eta}-1}{\gamma}\right)\right]\times \nonumber\\
\left[-i\hbar\frac{d}{d\eta}-
im\omega\left(\frac{e^{\gamma \eta}-1}{\gamma}\right)\right]\Bigg{\}}\phi(\eta)
\end{eqnarray}
which corresponds to the Klein-Gordon equation with the
non-minimal coupling $p_\gamma-
im\omega (e^{\gamma \eta}-1)/\gamma$
for relativistic
wave-function $\phi(\eta,t)=\psi(x(\eta),t)$ in the
$\eta$-space. It is worthing to note that Eqns.
\eqref{KG-Morse-oscillator-1D-1} and
\eqref{KG-Morse-oscillator-1D-2} extend the
equivalence between the harmonic oscillator with the generalized momentum $p_\gamma$ and the
Morse potential \cite{Costa_Filho-2013}, in the context of the KG equation.
Moreover,
 by redefining $\eta$ as $\widetilde{\eta}=\eta-\eta_0$ and $\phi(\eta)$ as $\widetilde{\phi}
 	(\widetilde{\eta})=
 	\phi
 	(\widetilde{\eta}+\eta_0)$
 the \eqref{KG-Morse-oscillator-1D-2} can be rewritten as
\begin{eqnarray}
\label{KG-Morse-oscillator-1D-3}
\mathcal{E}\widetilde{\phi}(\widetilde{\eta})=\left[-\frac{\hbar^2}{2m}\frac{d^2}{d\widetilde{\eta}^2}+\frac{m\widetilde{\omega}^2}{2\gamma^2}
(e^{\gamma\widetilde{\eta}}-1)^2-\frac{\hbar\widetilde{\omega}}{2}\right]\widetilde{\phi}(\widetilde{\eta})
\end{eqnarray}
where $\eta_0=
\frac{\ln(
	\widetilde{\omega}/\omega
	%\frac{\gamma^2\hbar}{2m\omega}
	)}{\gamma}$
and $\widetilde{\omega}=\omega
\left(1+\frac{\gamma^2\hbar}{2m\omega}\right)$
is a modified frequency with
$\eta_0$ a displacement of the origin of the potential, both arising due to the relativistic coupling.
By comparison between the
Eqns. \eqref{Morse-SE}--\eqref{eigenergies-QMO}
and the Eq.
\eqref{KG-Morse-oscillator-1D-3}
it is obtained the energy spectrum
of the KGMO (Klein-Gordon Morse oscillator)
\begin{eqnarray}\label{eigenergies-KGMO}
\mathcal{E}_N=\hbar\widetilde{\omega}
\left(N+\frac{1}{2}\right)\left[1-\frac{\gamma^2\hbar}
{2m\widetilde{\omega}}\left(N+\frac{1}{2}\right)\right]+\frac{\hbar\widetilde{\omega}}{2}
\end{eqnarray}
with $N=0,1,2,\ldots$. It is also
instructive
to obtain the null deformation ($\gamma\rightarrow0$) and
the non-relativistic ($\mathcal{E}_N\approx\epsilon_N$)
limits for the energy spectrum. In the former case,
from Eq.
\eqref{eigenergies-KGMO}
for $\gamma\rightarrow0$ $(\widetilde{\omega}\rightarrow\omega)$
we recover the
Eq. \eqref{KG-energies}
that corresponds to the energy levels of
Klein-Gordon oscillator except by an extra term
$\hbar\omega/2$, while in the later
case we obtain the QMO energies whose formula
is identical to the Eq.
\eqref{eigenergies-QMO} but with the
modified frequency $\widetilde{\omega}$.
In both cases the limits are not strictly
identical to the standard ones due to the non-minimal coupling employed.

For the one-dimensional Dirac Morse oscillator (DMO),
and using the representation $(\sigma_3,i\sigma_2)$ with the momentum $p_{\gamma}$ we obtain
\begin{eqnarray}
&(E-mc^2)\psi_1=
c(p_{\gamma}+
im\omega x)\psi_2 \nonumber \\
&(E+mc^2)\psi_2=
c(p_{\gamma}-
im\omega x)\psi_1, \nonumber
\end{eqnarray}
from which results
\begin{eqnarray}
\label{Dirac-oscillator-1D-1}
\mathcal{E}\psi_1=\frac{1}{2m}(p_{\gamma}+im\omega x)(p_{\gamma}-im\omega x)\psi_1,
\end{eqnarray}
that is identical to
\eqref{KG-Morse-oscillator-1D-1} for the spinor components ($\psi_1,\psi_2)$.
This is expected since a one-dimensional
particle cannot manifest
spin and angular momentum interactions,
which is reflected by the fact that $\mathbf{p_\gamma \times x}=0$.

In order to
validate
%test the validity of
this generalization, we will reproduce some of the S-wave states ($l=0$)
for the $\textrm{H}_2$ molecule,
with the parameters extracted from the Ref. \cite{H2}.
For accomplish this we make an adjustment on Eq. \eqref{KG-Morse-oscillator-1D-3}
expressed by the constants
$D_{\mathrm{e}}=\frac{m\widetilde{\omega}^2}{2\gamma^2}$, $\alpha =-\gamma r_{\mathrm{e}}$.
By replacing the values of $D_{\mathrm{e}}=4.7446$ eV, $r_{\mathrm{e}}=0.7416\r{A}$, $m=0.50391$ amu, $\alpha=1.440558$ and $E_0=\hbar^2/mr_{\mathrm{e}}^2=1.508343932$ eV in Eq. \eqref{KG-Morse-oscillator-1D-3} it is
obtained the eigenvalues equation
%($M$ denoting the Morse potential)
\begin{eqnarray}
\label{KG-Morse-oscillator-NASSER}
\mathcal{\lambda}\phi= \Bigg{\{}-\frac{\hbar^2}{2m}\frac{d^2}{d\eta^2}
+D_{\mathrm{e}}\left[e^{-2\alpha\left(\frac{\eta}{\eta_0}-1\right)}-2e^{-\alpha\left(\frac{\eta}{\eta_0}-1\right)}\right]^2\Bigg{\}}\phi
\end{eqnarray}
with the corresponding energies
$\mathcal{\lambda}_N=\mathcal{E}_N+\frac{\hbar\widetilde{\omega}}{2}-D_\textrm{e}$
\begin{eqnarray}\label{eigenergies-H2}
\mathcal{\lambda}_N=-\frac{\alpha^2E_0}{2}\left(\frac{2r_{\mathrm{e}}\sqrt{2mD_{\mathrm{e}}}}{\alpha\hbar}-\frac{1}{2}-N\right)^2.
\end{eqnarray}	
From the Table \ref{fig:table1} we see that the
non-relativistic energies
of the $\textrm{S}$-wave states
are a very good agreement with the literature
(see \cite{Nasser-2007} and references
therein), differing
only from the ninth decimal number.
\begin{table}[!htb]\label{fig:table1}
	\centering
	\tabcolsep=0.30cm
	\begin{tabular}{|c|c|c|}
		\hline
		&This work & \cite{Nasser-2007}
		\\
		\hline
		N=0&$4.476013136977448$ & $4.476013136943936$
		\\
		\hline
		N=1&$3.962315359052883$ & $3.962315358958284$
		\\
		\hline
		N=2&$3.479918845289036$ & $3.479918845141218$
		\\
		\hline
		N=3&$3.028823595685905$ & $3.028823595492864$
		\\
		\hline
	\end{tabular}
	\caption{Non-relativistic energies (in eV) of the
KGMO and DMO given by the formula \eqref{eigenergies-H2}
for some $S$-wave states ($l=0$ and $N=0,1,2,3$) of the $H_2$ molecule
along with
those obtained in Ref. \cite{Nasser-2007}.
The agreement is up to the eighth decimal number.}
\end{table}
\subsection{Three-dimensional case: the generic radial differential equation}
\begin{figure}[hbt]
\begin{minipage}[b]{1\linewidth}
\includegraphics[width=\linewidth]{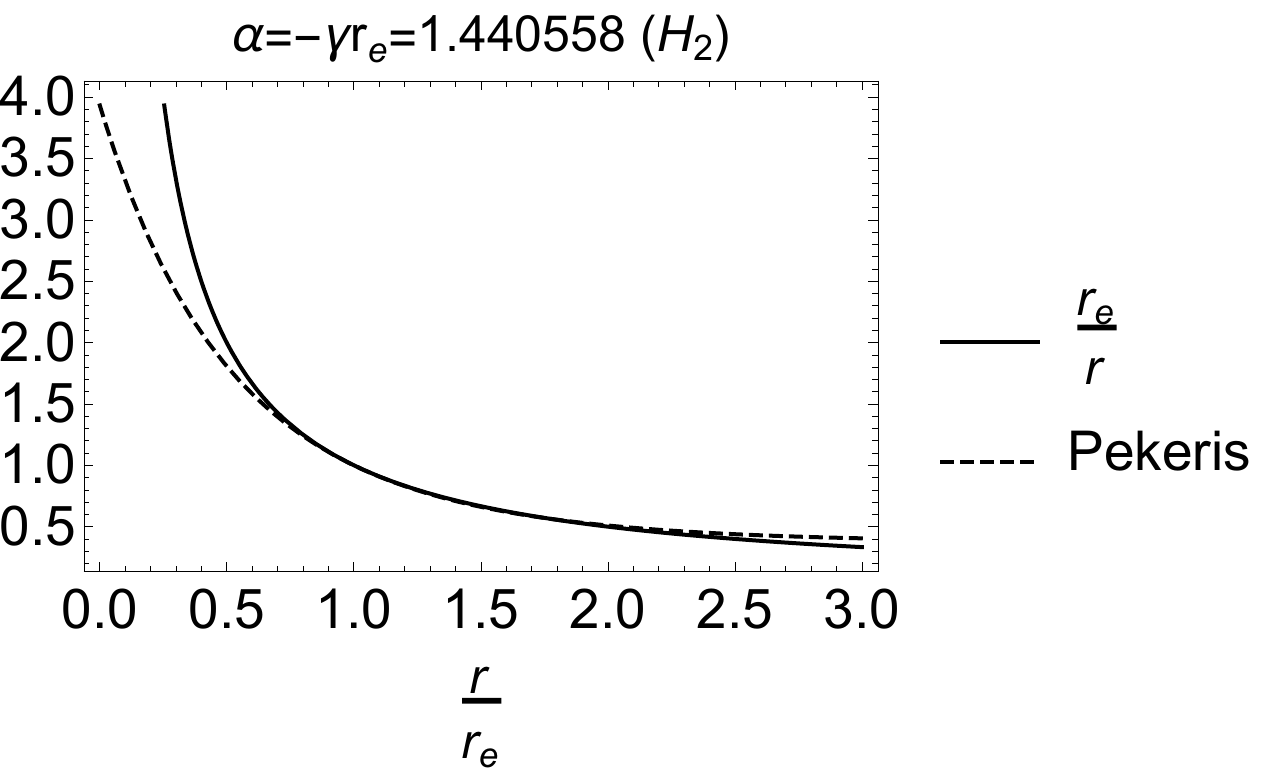}
\end{minipage}\\
\begin{minipage}[b]{1\linewidth}
\includegraphics[width=\linewidth]{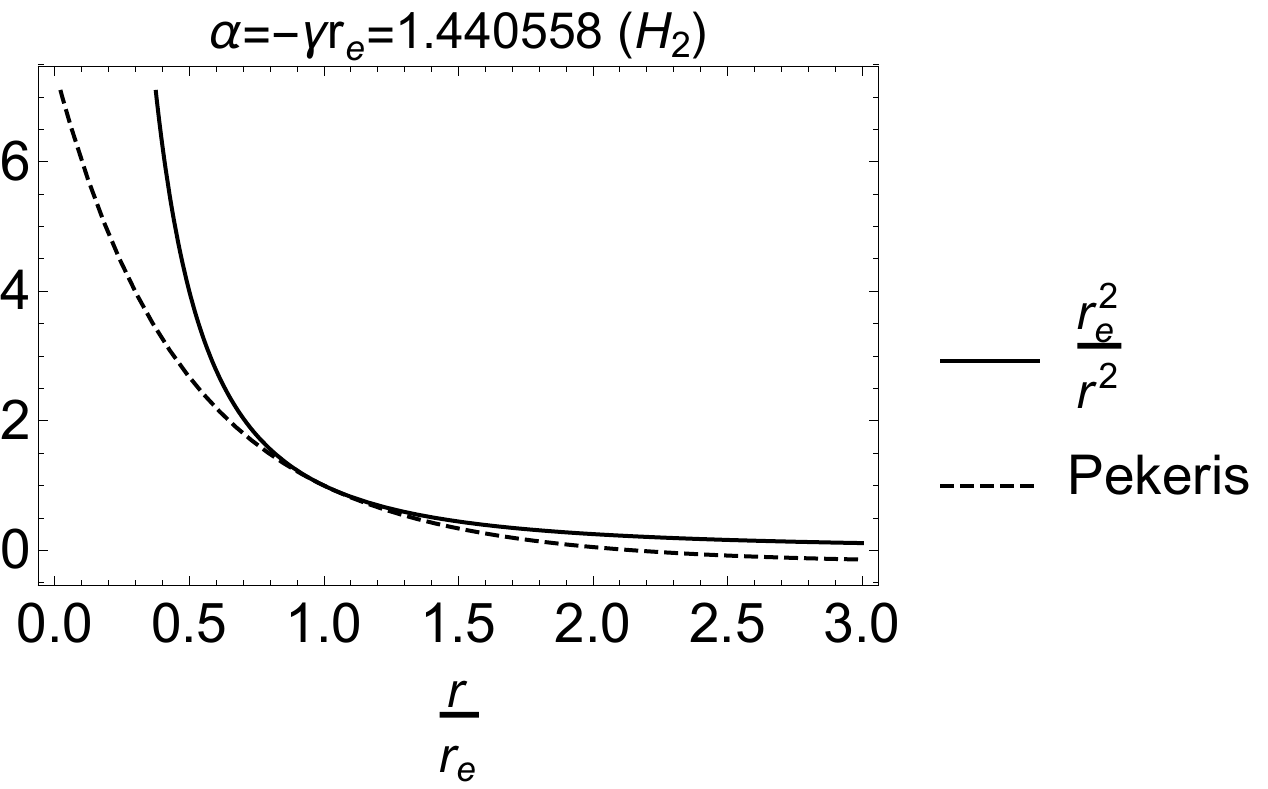}
\end{minipage}
\caption{\label{fig:fig1}
Accuracy of the generalized Pekeris approximation \eqref{generelized-Pekeris}
for the terms $r/r_e$ and $(r/r_e)^2$ in the case of the $\textrm{H}_2$ molecule with
Morse potential non-minimal coupling \eqref{deformed-potential}.
}
\end{figure}

The three-dimensional non-minimal couplings
lead to the radial differential equation \cite{Diracoulomb}
\begin{eqnarray}\label{structural-radial}
\mathcal{E}\Phi&=&\Bigg{\{}-\frac{\hbar^2}{2m}\frac{d^2}{dr^2}
+\frac{m\omega^2}{2}U^2-\frac{\hbar\omega}{2}\frac{dU}{dr}\nonumber\\
&&-[1+f(j,l)]\frac{\hbar\omega U}{r}+\frac{\hbar^2 l(l+1)}{2mr^2}\Bigg{\}}\Phi
\end{eqnarray}
where
\begin{eqnarray}\label{function-f}
f(j,l) &=& \begin{cases}
\begin{array}{lr}
0 \hspace{4cm}\text{for KG case;}\\
2[j(j+1)-l(l+1)-3/4] \hspace{0.1cm}\text{for Dirac case} \nonumber
\end{array}
\end{cases}
\end{eqnarray}
contains total spin and angular momentum effects.
The presence of the term $\propto U/r$ and the
centrifugal one $\propto 1/r^2$ makes \eqref{structural-radial}
to be not analytically solvable. The usual strategy for this case is to employ
the Pekeris approximation in both terms.
Now defining $U$ as a deformed linear potential
(recovering the harmonic potential
$U(r)=r-r_e$ when $\gamma\rightarrow0$) given by
\begin{eqnarray}\label{deformed-potential}
U(r)=\frac{e^{\gamma(r-r_{\mathrm{e}})}-1}{\gamma},
\end{eqnarray}
we can recast the generic radial equation as (see Appendix)
\begin{eqnarray}\label{Morse-radial-compact}
\mathcal{\widetilde{E}}\Phi(r)&=&\left(-\frac{\hbar^2}{2m}\frac{d^2}{dr^2}
+U_{\textrm{eff}}\right)\Phi(r),
\end{eqnarray}
where
$U_{\textrm{eff}}(r)=
\frac{m\Omega^2}{2\gamma^2}[e^{\gamma(r-r_{\textrm{eff}})}-1]^2+U_0$ can be considered
as a resultant effective Morse potential. In Fig. \ref{fig:fig1}
we illustrate the accuracy of the generalized Pekeris approximation
\eqref{generelized-Pekeris} compared with the
coulomb and the centrifugal terms, $r/r_e$ and $(r/r_e)^2$,
for the $\textrm{H}_2$ molecule with $\alpha=-\gamma r_e=1.440558$.
It can be seen that the Pekeris approximation fits well both terms within
the interval $r/r_e\in(0.5,1.5)$, which justifies its employment
for the vibrational states $r\sim r_e$.

\begin{figure}[hbt]
\begin{minipage}[b]{0.9\linewidth}
\includegraphics[width=\linewidth]{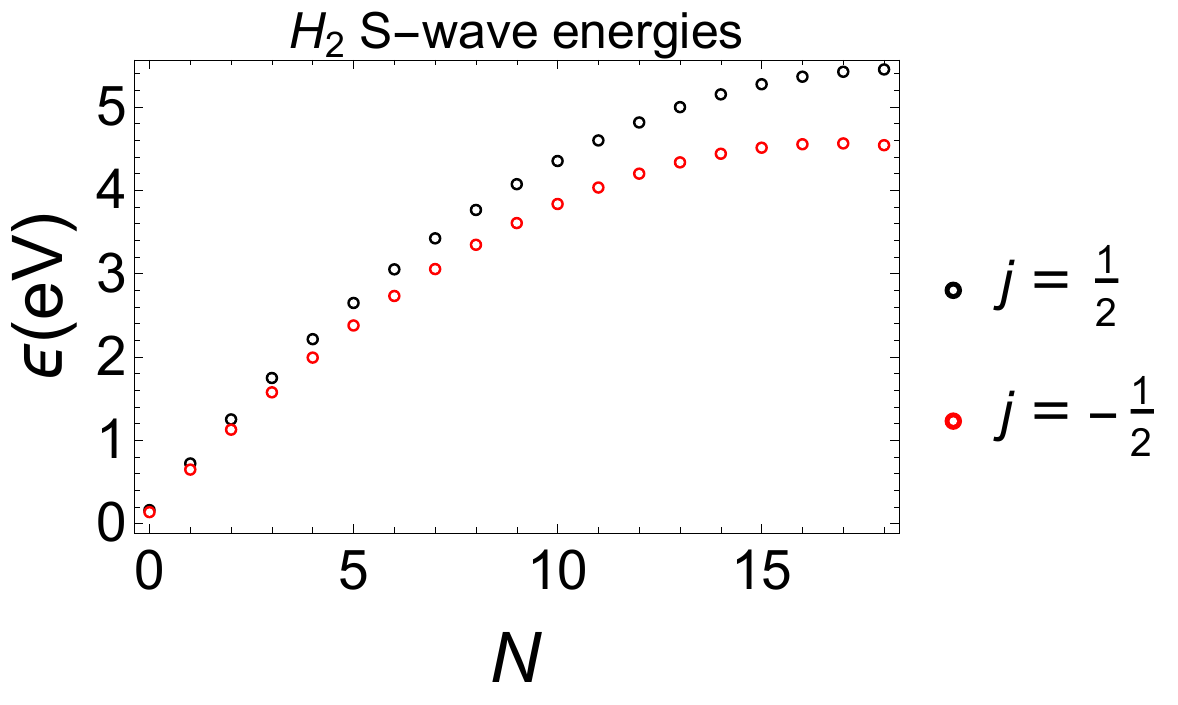}
\end{minipage}\\
\begin{minipage}[b]{0.9\linewidth}
\includegraphics[width=\linewidth]{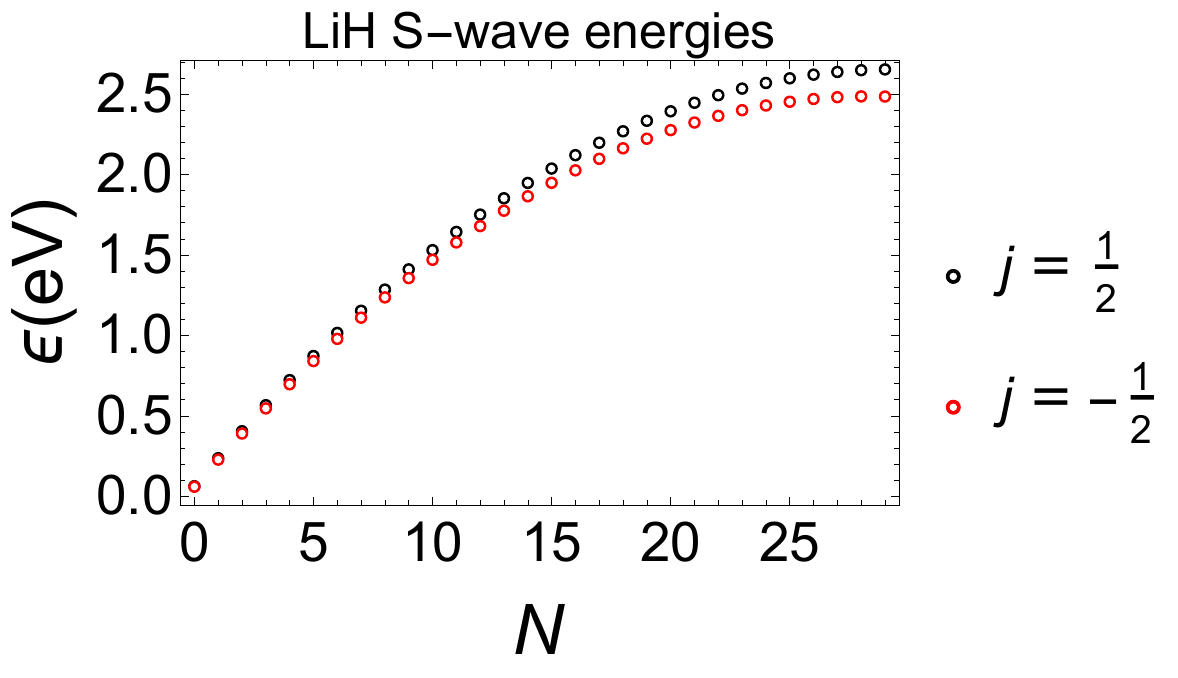}
\end{minipage}
\caption{\label{fig:fig2}
Energies \eqref{eq:energy-recasted} in eV
of the $\textrm{H}_2$ and the LiH molecules of the allowed S-wave states ($l=0$).
In both cases the projection spin $j=+1/2$ increases the energy and the states hold non-degenerated within
the ranges $0\leq N\leq18$ ($\textrm{H}_2$) and $0\leq N\leq 29$ (LiH).}
\end{figure}

We obtain the energies of the KGMO and DMO in the Pekeris approximation, given by
\begin{eqnarray}\label{eigenergies-KG-Dirac-Morse}
\mathcal{\widetilde{E}}_N=\hbar\Omega\left(N+\frac{1}{2}\right)\left[1-\frac
{\gamma^2\hbar}{2m\Omega}\left(N+\frac{1}{2}\right)\right]
+U_0,
\end{eqnarray}
which for
$\gamma\rightarrow0$ allows to recover
a relativistic harmonic oscillator-like energy
\begin{eqnarray}\label{eigenergies-KG-Dirac-Morse-oscillator-limit}
\mathcal{\widetilde{E}}_N\longrightarrow \hbar\Omega\left(N+\frac{1}{2}\right)+U_0
\end{eqnarray}
where the modified frequency is (see Appendix)
\begin{eqnarray}\label{modified-frequency}
\Omega^2\longrightarrow
\omega^2A(\alpha,\delta,j,l)\left[1-B\left(\alpha,\delta,j,l)/(2A(\alpha,\delta,j,l)\right)\right]^2
\end{eqnarray}
with
\begin{eqnarray}\label{modified-U0}
U_0=\frac{m\Omega^2}{2\gamma^2}\left[C(\alpha,l)-B(\alpha,\delta,j,l)^2/(4A(\alpha,\delta,j,l)^2)\right].
\end{eqnarray}
The functions $A(\alpha,\delta,j,l), B(\alpha,\delta,j,l)$ and $C(\alpha,l)$
express the generalized Pekeris approximation \eqref{generelized-Pekeris}
in terms of the angular momentum $l$ and the total
spin $j$ (by means of
$f(j,l)$) along with the parameterization $\alpha=-\gamma r_e>0$ and $\delta=\frac{\hbar}{m\omega r_e^2}>0$.
The differences between the Dirac oscillator energies \eqref{Dirac-energies}
and its corresponding limit case
of null deformation of the DMO given by
\eqref{eigenergies-KG-Dirac-Morse-oscillator-limit}
manifest that the Pekeris approximation do not allow to make
a perfect limit
but rather
to obtain effective oscillator.
Considering some diatomic molecules it can be seen that
$\delta\ll1$ represents a typical situation, which is shown in
Table \ref{fig:table2} along with
the $N_{\textrm{max}}$ of allowed S-wave states ($l=0$).
\begin{table}[!htb]\label{fig:table2}
	\centering
	\tabcolsep=0.30cm
	\begin{tabular}{|c|c|c|c|}
		\hline
		Molecule & $\delta$ & $N_{\textrm{max}}$ & $D_\textrm{e}$ (eV) \cite{Nasser-2007}
		\\
		\hline
		$\textrm{H}_2$ &$0.0276729$ & $18$ & $4.7446$
		\\
		\hline
		$\textrm{LiH}$ &$0.0106979$ & $29$ & $2.515287$
		\\
		\hline
		$\textrm{HCl}$ &$0.00708095$ & $24$ & $4.61907$
		\\
		\hline
		$\textrm{CO}$ &$0.00177962$ & $83$ & $11.2256$
		\\
		\hline
	\end{tabular}
	\caption{Parameter $\delta=\sqrt{E_0/(2\alpha^2 D_\textrm{e})}$ in function of the
    Hartree energy $E_0=\frac{\hbar^2}{mr_e^2}$ and the dissociation energy $D_\textrm{e}$, along with
    the maximum number of S-wave states ($l=0$) that are allowed in the
    generic equation \eqref{Morse-radial-compact}.}
\end{table}
In this regime
%the energy gap results vanishingly small compared with the oscillator potential energy, expressed equivalently by
%$m\omega^2 r_e^2\gg \hbar\omega$, and
we can interpret the effects
of the spin and angular momentum contributions to the energy as follows.
When $\delta\ll1$ all the quadratic terms $\propto\delta^2$
in $A,B$ and $C$ can neglected so $\Omega^2$ turns out $\approx$
$\omega^2\left(1+\delta\left(3+2(1+f(j,l))\frac{1+\alpha}{\alpha}+l(l+1)\frac{\alpha^2+3\alpha+2}{\alpha}\right)\right)$, thus
carrying all the differences respect to the one-dimensional case
(along with the spin and angular momentum contributions) in the term with $\delta$.
Also, the term $U_0$ results $\approx$ $\frac{m\omega^2}{2\gamma^2}\delta(l(l+1)-1)$.
In Fig. \ref{fig:fig2} it is shown the allowed energies
of the $\textrm{H}_2$ and the LiH molecules of the S-wave states.

\section{Thermodynamics of the S-wave states and Morse spectrum finiteness:
Schottky effect}
We explore the effects of the approximated radial equation
\eqref{Morse-radial-compact}
in the statistical properties of
the S-wave states ($l=0$) that manifest the vibrational features of
the system.
With the aim to obtain the partition function
of the canonical ensemble, we consider that the system is in equilibrium with a
thermal bath of finite temperature $T$. We shall consider only the states
with positive energy to avoid the negative energies that are unlimited by below, which
also guarantees a stable ensemble \cite{Pacheco-EPL}.
For reasons of calculus we recast the formula of the energy
\begin{figure}[hbt]
\begin{minipage}[b]{0.7\linewidth}
\includegraphics[width=\linewidth]{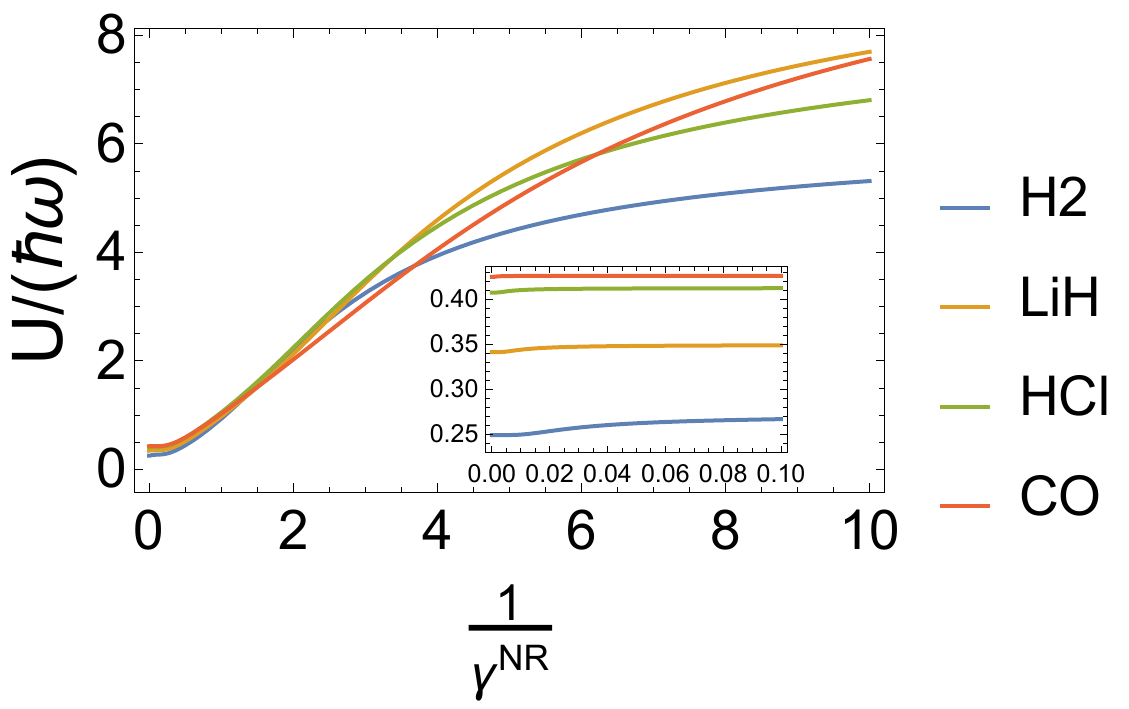}
\end{minipage}\\
\begin{minipage}[b]{0.7\linewidth}
\includegraphics[width=\linewidth]{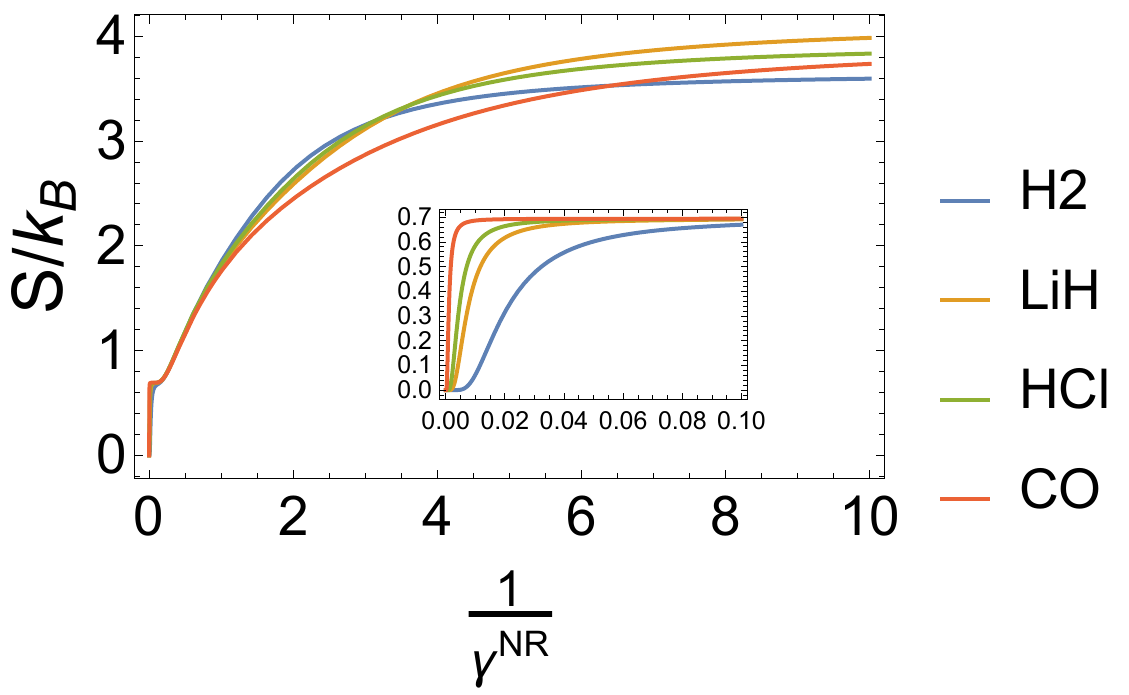}
\end{minipage}\\
\begin{minipage}[b]{0.7\linewidth}
\includegraphics[width=\linewidth]{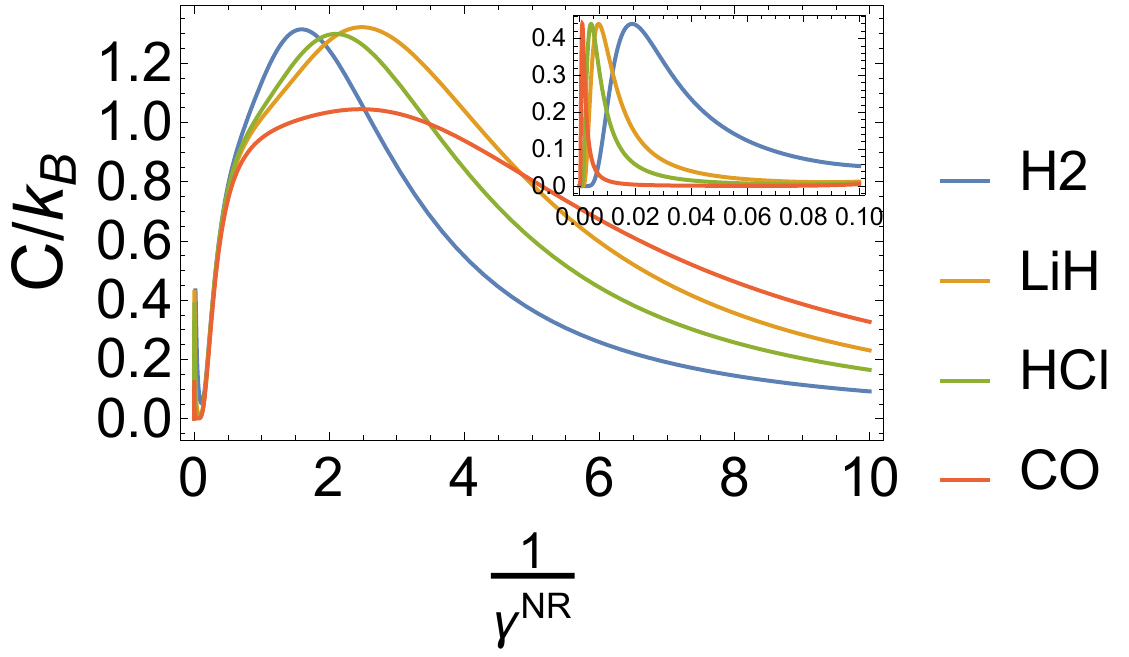}
\end{minipage}
\caption{\label{fig:fig3}
(a) Internal energy (top), (b) entropy (center) and (d) heat capacity (bottom)
of the S-wave states ($l=0$) of the $\textrm{H}_2$, LiH, HCl and CO molecules in the
non-relativistic regime in the presence of the effective radial Morse potential
\eqref{Morse-radial-compact} for the Dirac case. The parameters of the Table \ref{fig:table2} along with
the approximated energy \eqref{eq:energy-recasted} (for $\delta^2$ vanishingly small) were employed.
Two peaks are observed for the heat capacity (the first one in the inset),
in virtue of the finiteness of the energy levels (Schottky effect).}
\end{figure}
\eqref{eigenergies-KG-Dirac-Morse} as
\begin{eqnarray}\label{eq:energy-recasted}
&\mathcal{\widetilde{E}}_{N,j=\pm1/2}/\hbar\omega=
\sqrt{1+\delta(3\pm 2(1+\alpha)/\alpha)}\left(N+\frac{1}{2}\right)\times \nonumber\\
&\bigg[1-\frac{\alpha^2\delta}{2\sqrt{1+\delta(3\pm 2(1+\alpha)/\alpha)}}\left(N+\frac{1}{2}\right)\bigg]-
\frac{1}{2\alpha^2}
\end{eqnarray}
where $\pm$ stands for the spin projections $+1/2$ and $-1/2$ respectively.
Thus, we can perform two partitions functions
\begin{equation}\label{eq:partition-function-nonrelativistic}
Z^{\textrm{NR}}=\sum_{j=\pm1/2}\sum_{N=0}^{N_{\textrm{max}}}e^{\gamma^\textrm{NR}\mathcal{\widetilde{E}}_{N,j=\pm1/2}/\hbar\omega}
\end{equation}
and
\begin{equation}\label{eq:partition-function-relativistic}
Z^{\textrm{R}}=\sum_{j=\pm1/2}
\sum_{N=0}^{N_{\textrm{max}}}e^{\gamma^\textrm{R}\sqrt{1+2\gamma\mathcal{\widetilde{E}}_{N,j=\pm1/2}/\hbar\omega}}
\end{equation}
\begin{figure}[hbt]
\begin{minipage}[b]{0.7\linewidth}
\includegraphics[width=\linewidth]{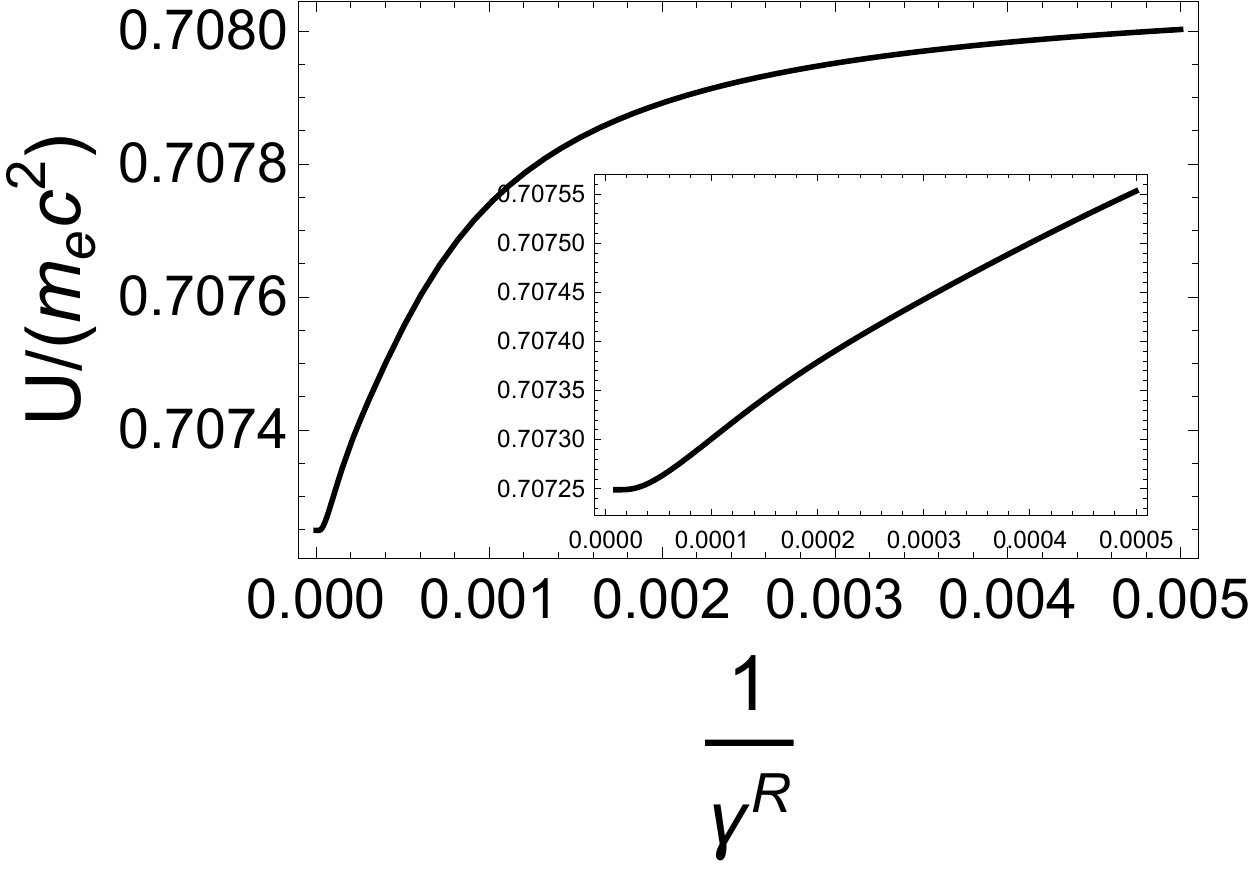}
\end{minipage}\\
\begin{minipage}[b]{0.7\linewidth}
\includegraphics[width=\linewidth]{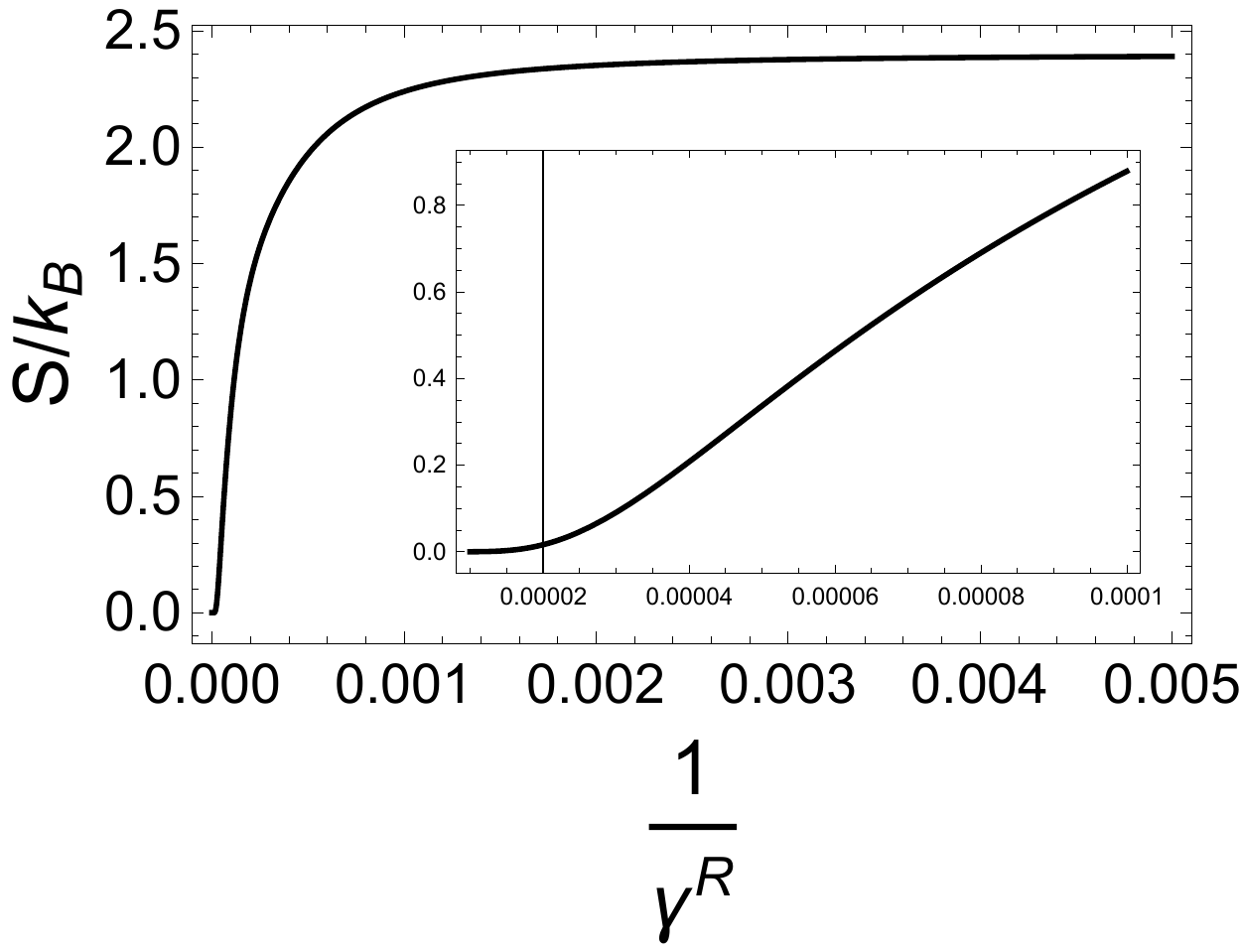}
\end{minipage}\\
\begin{minipage}[b]{0.7\linewidth}
\includegraphics[width=\linewidth]{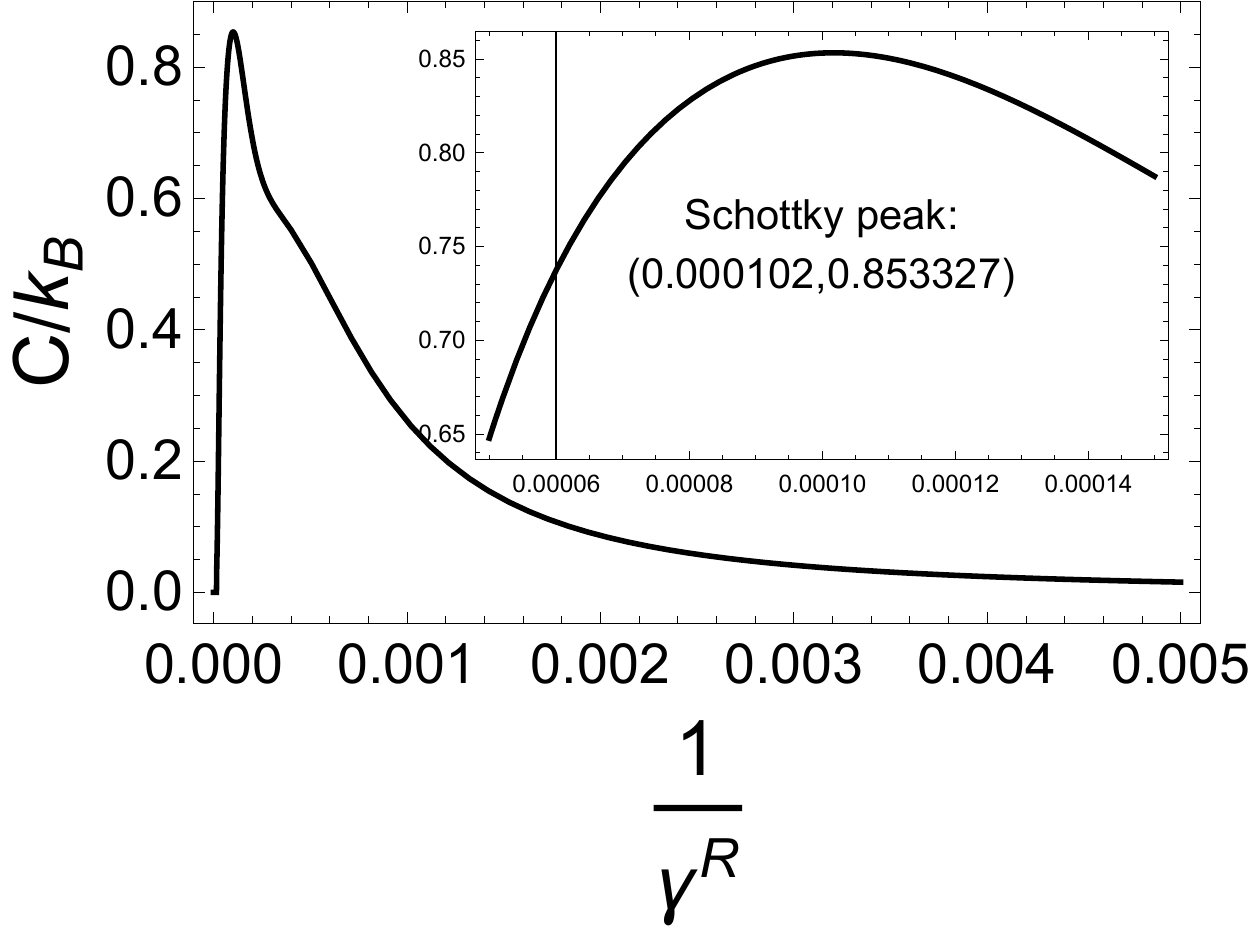}
\end{minipage}
\caption{\label{fig:fig4}
(a) Internal energy (top), (b) entropy (center) and (d) heat capacity (bottom)
of the S-wave states ($l=0$) of electrons with $\hbar\omega\sim 124 \textrm{eV}$ (ultraviolet spectrum) in
the relativistic regime in the presence of the effective radial Morse potential
\eqref{Morse-radial-compact} for the Dirac case.
The parameters
$\gamma=0.000242661=\hbar\omega/m_ec^2$, $\delta=E_{\textrm{h}}/\hbar\omega=0.219444$ and
$\alpha=1$ along with
the approximated energy \eqref{eq:energy-recasted} were employed.
As in the molecules case, a peak is observed for the heat capacity due to the Schottky effect.}
\end{figure}
corresponding to the non-relativistic and the relativistic cases. Here the
dimensionless parameters $\gamma^\textrm{NR}=\frac{\hbar\omega}{k_BT}$,
$\gamma^\textrm{R}=\frac{mc^2}{k_BT}$ measure the ratios between the
vibrational energy and the rest mass energy with respect to the thermal excitations, and
$\gamma=\frac{\hbar\omega}{mc^2}$ measures the ratio between the vibrational energy and the rest mass one.
These coefficients allow to characterize all the regimes of interest from the low to the high temperatures as well as
the intermediate ones. It is also assumed that $\gamma$ is fixed for each molecule of the Table \ref{fig:table2}
in terms of its characteristic parameters. To complete our analysis the (dimensionless) thermodynamical potentials
are needed
\bey\label{thermodynamical-relations}
U &=& -\frac{\partial \ln Z^\textrm{NR,R}}{\partial\gamma^\textrm{NR,R}} \quad \textrm{(internal energy)} \nonumber\\
F &=& -\frac{1}{\gamma^\textrm{NR,R}}\ln Z^\textrm{NR,R}  \quad \textrm{(Helmholtz free energy)} \nonumber\\
S &=& (\gamma^\textrm{NR,R})^2\frac{\partial F}{\partial \gamma^\textrm{NR,R}}  \quad \textrm{(entropy)} \nonumber\\
C &=& -(\gamma^\textrm{NR,R})^2\frac{\partial U}{\partial \gamma^\textrm{NR,R}}  \quad \textrm{(heat capacity)}
\eey
from which all the thermodynamics of the S-wave states can be derived. For recovering
the units of the thermodynamical potentials it is enough to add the
energy factor $\hbar\omega$ or $mc^2$ in $U$ and $F$, and to add $k_B$ in $S$ and $C$.
The notations $Z^\textrm{NR,R}$ and $\gamma^\textrm{NR,R}$ stand for their respective magnitudes
in the non-relativistic and relativistic contexts.
For the molecules above mentioned the coefficient $\gamma=\hbar\omega/mc^2$ results vanishingly small,
due to their enormous value of the rest mass $mc^2$ (of the order of the $\sim 1000$ Mev)
against the small photon energy $\hbar\omega$ characteristic of the
level spacements in typical quantum transitions. So, in order to
see relativistic effects and to maintain $\delta\ll1$ we shall consider $\hbar\omega\sim 124 \textrm{eV}$, that corresponds
to the extreme ultraviolet spectrum, along with $mc^2=m_ec^2=511 \textrm{keV}$ (i.e. the rest mass of the electron).
In virtue that $\delta=\textrm{E}_{\textrm{h}}/\hbar\omega$ with $\textrm{E}_{\textrm{h}}$ the Hartree energy,
we have $\gamma=0.000242661$ and $\delta=0.219444$ so $\delta^2=0.0481555$ can be neglected in relation with $\delta$, and then
the approximation \eqref{eq:energy-recasted} holds valid. We set $\alpha=-\gamma r_e=1$.
Thus, the number $N_\textrm{max}$ of allowed states for the states
with projection spin $1/2$ and $-1/2$ result $6$ and $3$ respectively
\footnote{As in the case of the Table 2,
the maximum number of allowed is calculated from $\alpha$ and
$\delta$ by the formula $N_\textrm{max}=[(\delta\alpha^2)^{-1}\sqrt{1+\delta(3\pm2(1+\alpha)/\alpha)}-1/2]$, with
$\pm$ and
$[x]$ standing for the spin projection and the integer part of $x$.}.

From Figs. \ref{fig:fig3} and \ref{fig:fig4}
we see the Schottky effect is present in both regimes, the non-relativistic and the relativistic one, due to the finite
number of allowed states. The peaks in the heat capacity are physically interpreted
due the fact that the more higher is the temperature the less number of states that the system has to be possibly occupied.
So when the temperature sufficiently increases that the factor
$k_BT$ approaches to the difference of the energy levels, a peak in the heat capacity emerges, and from there
small changes in the temperature produce changes in the entropy in such a way the heat capacity
continues decreasing up to be zero for $T\rightarrow\infty$.
For comparing the Dirac and KG cases, the behavior of the Schottky peaks and their
critical temperatures for the systems studied is shown in Table 3.
\begin{table}
[!htb]
\label{fig:table3}
	\centering
	\tabcolsep=0.1cm
	\begin{tabular}{|c|c|c|c|c|}
		\hline
		System & $C(T_c)$ (Dirac) & $C(T_c)$ (KG) & $T_c$ (Dirac)  & $T_c$ (KG)
		\\
		\hline
		$\textrm{H}_2$ & $0.44$ & $1.8\times10^{-11}$ & $-154.1$ & $-265.2$
		\\
                       & $1.31$ & $1.29$ & $9804.4$ & $10606.6$
        \\
		\hline
		$\textrm{LiH}$ & $0.44$ & $7.474\times10^{-11}$ & $-259.2$ & $-271.7$
		\\
                       & $1.32$ & $1.31$ & $4749.6$ & $4876.6$
        \\
		\hline
		$\textrm{HCl}$ &$0.44$ & $1.34\times10^{-11}$ & $-255.1$ & $-270.$
	     \\
                       & $1.3$ & $1.29$ & $8750.6$ & $8883$
        \\
		\hline
		$\textrm{CO}$ &$0.44$ & $8.99\times10^{-11}$ & $-269.8$ & $-270.5$
        \\
                       & $1.04$ & $1.04$ & $7435.2$ & $7454.71$
        \\
		\hline
		$e^{-}$ &$0.85$ & $1.11$ & $6.03\times10^{5}$ & $2.11\times10^{6}$
		\\
		\hline
	\end{tabular}
	\caption{Schottky peaks (in units of $k_B$) of the systems studied and their associated critical temperatures in Celsius for the
    Dirac and KG cases. The differences between the Dirac and KG cases are appreciable in the low temperature limit (first row of each molecule)
    or in the relativistic
    regime (electron case).}
\end{table}
We can see that for the molecules the first peaks give place at low temperatures within the range
of the $-270\leq T\leq-150$ degrees Celsius, while the second peaks arise in the interval $4700\leq T\leq10700$
that correspond to thermal energies $k_BT$
of the order of the dissociation energy of the molecules $\sim$ eV. For temperatures $T\geq10700$
the predictions of the Morse model are no longer valid
and then the heat capacity exhibits a typical decreasing with the temperature.
In the non-relativistic regime of the molecules studied the
spin effects are predominant at the low temperature regime $[-270,-150]$, with an appreciable difference
in the magnitude order of the
values of the peaks of the Dirac and KG cases.
On the other hand, for the electron provided with a high energy $\hbar\omega\sim 124$ eV the spin contributions to the heat capacity
are visible still
at high temperatures $\sim10^5-10^6$.

\section{
General Pekeris approximation revisited: mapping from
three-dimensional radial equation to one-dimensional
Schr\"odinger-like equation}

In order to solve the generic radial equation \eqref{structural-radial}
with an arbitrary spherical potential $U(r)$,
%we show that it is possible to extend the generalized Pekeris approximation \eqref{generelized-Pekeris}
%obtained for the Morse potential. We
we revisit the generalized Pekeris approximation
of \cite{Pekeris2} by establishing the variable of the Pekeris expansion $y$ in function of the
potential coupling $U(r)$, and by deducing the family of potentials from which a mapping
onto a Schr\"odinger equation with non-minimal coupling emerges.
More generally, if $U(r)$ represents an spherical radial potential, we can define the dimensionless variable
\beq\label{generic-y}
y=\gamma U(r)+1=f^{-1}(\gamma(r-r_e))
\eeq
with $\gamma$ a real parameter having units of distance$^{-1}$,
$f^{-1}(x)=\gamma U(x/\gamma+r_e)+1$ and
$f(x)=\gamma U^{-1}((x-1)/\gamma)-\gamma r_e$.
In particular, for the
Morse potential coupling $U(r)=(1/\gamma)(e^{\gamma(r-r_e)}-1)$ we recover the previously used
$y=e^{\gamma(r-r_e)}$ with $f^{-1}(x)=e^x$. In order to provide the method, we assume that $U$ has a
differentiable inverse
$U^{-1}$ and then $f$ and $f^{-1}$ result also differentiable. Thus, from \eqref{generic-y} it follows
approximated expressions for
$r_e/r$ and $(r_e/r)^2$ up to terms of order $2$ around $y=1$ ($r=r_e$)
\bey\label{pre-general-pekeris}
r_e/r&=&\left(1+f(y)/(\gamma r_e)\right)^{-1}\approx \frac{1}{\frac{f(1)}{\gamma  r_e}+1}+\sum_{i=1}^{2}a_i(y-1)^i\nonumber\\
(r_e/r)^2&=&\left(1+f(y)/(\gamma r_e)\right)^{-2}\approx \frac{1}{\left(\frac{f(1)}{\gamma  r_e}+1\right)^2}+\sum_{j=1}^{2}a_j(y-1)^j \nonumber\\
a_1&=&-\frac{\gamma  r_e f'(1)}{\left(\gamma  r_e+f(1)\right)^2}\nonumber\\
a_2&=&-\frac{\left(\gamma  r_e \left(\gamma  r_e f''(1)+f(1) f''(1)-2 f'(1)^2\right)\right)}{2 \left(\gamma  r_e+f(1)\right)^3}\nonumber\\
b_1&=&-\frac{2 (y-1) \left(\gamma ^2 r_e^2 f'(1)\right)}{\left(\gamma  r_e+f(1)\right)^3}\nonumber\\
b_2&=&-\frac{\gamma ^2 r_e^2 \left(\gamma  r_e f''(1)+f(1) f''(1)-3 f'(1)^2\right)}{\left(\gamma  r_e+f(1)\right)^4},
\eey
which for the Morse potential case result more simplified since $f(1)=0$.
%where we used that $f(1)=0$, $f(x)=\gamma U^{-1}((x-1)/\gamma)-\gamma r_e$ and
%$U(r_e)=0$.
For avoiding terms of the type
$\propto U^3$ in the term $U/r$ in \eqref{potential-radial} we can still make $a_2=0$.
Then, using \eqref{pre-general-pekeris} the effective potential
\eqref{potential-radial} can be recasted
in terms of $y-1=\gamma U$ for $r\sim r_e$ as
\bey\label{effective-potential-mapping}
&U_{\mathrm{eff}}(r)\approx
%\frac{m\omega^2}{2\gamma^2}(y-1)^2-\frac{\hbar\omega}{2\gamma r_e}\frac{d(y-1)}{dr}
%-[1+f(j,l)]\frac{\hbar\omega}{\gamma r_e}\bigg[(y-1)\nonumber\\
%&+a_1(y-1)^2\bigg]+\frac{\hbar^2 l(l+1)}{2r_e^2}\left(1+b_1(y-1)+b_2(y-2)^2\right)=\nonumber\\
\left(\frac{m\omega^2}{2\gamma^2}-[1+f(j,l)]\frac{\hbar\omega}{\gamma r_e}a_1+\frac{\hbar^2 l(l+1)}{2r_e^2}b_2\right)(y-1)^2\nonumber\\
&+\left(-[1+f(j,l)]\frac{\hbar\omega}{\gamma r_e}+\frac{\hbar^2 l(l+1)}{2r_e^2}b_1\right)(y-1)
+\frac{\hbar^2 l(l+1)}{2r_e^2}-\frac{\hbar\omega}{2\gamma}\frac{d(y-1)}{dr}\nonumber\\
&=A_1\gamma^2 \bigg[U(r)+A_2/(2\gamma A_1)\bigg]^2+A_3-A_2^2/4A_1-\frac{\hbar\omega}{2}\frac{U(r)}{dr}
\eey
with the pertinent identifications for the constants $A_1,A_2,A_3$.
Hence, we arrive to one of the main results of the paper.
For $r\sim r_e$ we can map the differential radial equation \eqref{structural-radial} for an arbitrary spherical
potential $U(r)$ such that $U$ and its inverse $U^{-1}$ are differentiable in a neighbouring of $r=r_e$ and
of $U(r_e)$ respectively, into the one-dimensional Schr\"odinger-like equation
\bey\label{structural-mapping}
\mathcal{E}\Phi&=&\Bigg{\{}-\frac{\hbar^2}{2m}\frac{d^2}{dr^2}+A_1\gamma^2
\bigg[U(r)+A_2/(2\gamma A_1)\bigg]^2\nonumber\\
&&+A_3-A_2^2/4A_1-\frac{\hbar\omega}{2}\frac{U(r)}{dr}\Bigg{\}}\Phi,
\eey
that does not contain cross terms of the type $U(r)/r$ nor Coulomb or centrifugal terms
as the radial equation \eqref{structural-radial}.
The constants $A_1,A_2,A_3$ are determined by
\bey\label{constants-mapping}
A_1&=&\left(\frac{m\omega^2}{2\gamma^2}-[1+f(j,l)]\frac{\hbar\omega}{\gamma r_e}a_1+\frac{\hbar^2 l(l+1)}{2r_e^2}b_2\right) \nonumber\\
A_2&=&\left(-[1+f(j,l)]\frac{\hbar\omega}{\gamma r_e}+\frac{\hbar^2 l(l+1)}{2r_e^2}b_1\right) \nonumber\\
A_3&=&\frac{\hbar^2 l(l+1)}{2r_e^2},
\eey
which together with \eqref{pre-general-pekeris} and $f(x)=\gamma U^{-1}((x-1)/\gamma)-\gamma r_e$
give a complete proof of the desired mapping.
To emphasize its construction, we refer to the formula \eqref{structural-mapping}
as a \emph{Pekeris mapping}. Besides the Morse potential previously studied,
next we shall examine other illustrative examples.

\subsection{Example 1: modified 12-6-9 Lennard-Jonnes potential}

A classical example for
modelling the intermolecular interactions between a pair of neutral atoms of molecules
is the Lennard-Jones potential, that we can consider in a modified 12-6-9 form\footnote{12-6-9
refers to the sequence of the exponents in the terms $(r_e/r)^{n}$.}
\bey\label{Lennard-Jonnes-potential}
V(r)=\epsilon \bigg[\left(\frac{r_e}{r}\right)^{6} - \sqrt{2}\left(\frac{r_e}{r}\right)^3\bigg]^2=
V_{LJ}-2\epsilon \sqrt{2}\left(\frac{r_e}{r}\right)^{9}
\eey
with $\epsilon$ the depth of the potential well and $V_{LJ}$ standing for the Lennard-Jonnes potential.
In this case, from the identifications $\epsilon=m\omega^2/2\gamma^2$ and $\gamma=1/r_e$ we deduce
the non-minimal coupling $U(r)=(1/\gamma)[(\gamma r)^{-6}-\sqrt{2}(\gamma r)^{-3}]$,
and then by inverting $U(r)$ (where we choose the positive branch) it follows $f(x)$
\bey\label{f-Lennard-Jonnes}
f(x)=\left(\frac{2}{\sqrt{2}+\sqrt{2+4(x-1)}}\right)^{1/3}-1,
\eey
where $x=1$ corresponds to $U^{-1}(0)$, i.e. the minimum of the potential $r=r_e/2^{1/6}$.
Having obtained $f(x)$ the Pekeris mapping follows straightforwardly by
\eqref{pre-general-pekeris}, \eqref{structural-mapping} and \eqref{constants-mapping}.

\subsection{Example 2: homographic-squared potential}

Other type of invertible non-minimal coupling $U(r)$ that we can consider is an
homographic-squared potential, expressed by
\beq\label{homographic-potential}
V(r)=\epsilon\left(\frac{a(r/r_e)+b}{c(r/r_e)+d}\right)^2
\eeq
which for $c=0,d=1$ and $a=0$; $b,c\neq 0$ collapse in the
harmonic and the $1/r^2$ potentials respectively. Again,
$\epsilon$ represents the strength force of the potential and we can make the identifications
$\epsilon=m\omega^2/2\gamma^2$ and $\gamma=1/r_e$. Then, the non-minimal coupling results
$U(r)=\frac{1}{\gamma}(a(\gamma r)+b)/(c(\gamma r)+d)$. By inverting the
homographic coupling $U(r)$ we obtain
\bey\label{f-homographic}
f(x)=\frac{-b(x-1)+d}{-c(x-1)+a}-1.
\eey
In this case, the choice $f(1)=0$ simplifies the constants $a_1,b_1,b_2$ of
\eqref{pre-general-pekeris} and also implies
$a=d$.

\subsection{Example 3: Pekeris mapping into Schr\"odinger equation with non-minimal coupling }

We also can consider the special family of non-minimal couplings $U(r)$ satisfying the differential equation
\beq\label{family-potential}
\frac{dU}{dr}=\alpha_2 U^2 + \alpha_1 U + \alpha_0
\eeq
with $\alpha_i$ ($i=0,1,2$) real coefficients.
We notice that
%the non-minimal couplings $U(r)$ satisfying
\eqref{family-potential}
and the expressions \eqref{structural-mapping} allow to
rewrite the radial equation \eqref{structural-radial} with the effective potential being a
quadratic function of $U$, i.e.
\bey\label{radial-exact}
\mathcal{E}\Phi&=&\left[-\frac{\hbar^2}{2m}\frac{d^2}{dr^2}
+K_1(U(r)-K_2)^2+K_3\right]\Phi,
\eey
where $K_1,K_2,K_3$ are constants to be determined (with the help of \eqref{pre-general-pekeris})
and that depend on the quantum numbers $n,j,l$ along with
the parameters $m,\gamma,r_e,\omega,\alpha_1,\alpha_2,\alpha_3$.
In this case we say that the Pekeris mapping is allows to rewrite
\eqref{radial-exact} as
\bey\label{mapping-exact}
2m\mathcal{E}\Phi&=&[\left(p_r-iK_1(U(r)-K_2)\right)\left(p_r-iK_1(U(r)-K_2)\right)^{\dagger}\nonumber\\
&&+K_3]\Phi.
\eey
that is the free one-dimensional Schr\"odinger equation
dotted with the non-minimal coupling $p_r-iK_1(U(r)-K_2)$ in the $r$-direction.
By solving \eqref{family-potential} we find out what are the potentials that
belong to the Pekeris mapping \eqref{radial-exact}, which are given by
\bey\label{solutions-exact-mapping}
&&U(r)=\sqrt{4 \alpha _0 \alpha _2-\alpha _1^2} \times\nonumber\\
&&\frac{\tan \left[\frac{1}{2} \left(\sqrt{4 \alpha _0 \alpha _2-\alpha _1^2} K+
\sqrt{4 \alpha _0 \alpha _2-\alpha _1^2} r\right)\right]-\alpha _1}{2 \alpha _2}
\eey
with $K$ an arbitrary integration constant. Some representative
non-minimal couplings generated by the family
\eqref{solutions-exact-mapping} are shown in Table 4.
\begin{table}
[!htb]
\label{fig:table3}
	\centering
	\tabcolsep=0.1cm
	\begin{tabular}{|c|c|c|c|c|}
		\hline
		coupling type & $\alpha_2$ & $\alpha_1$ & $\alpha_0$ & $U(r)$
		\\
		\hline
		tangent & $1$ & $0$ & $1$ & $\tan\left(r+K\right)$
		\\
        \hline
        Morse & $<0$ & $0$ & $\alpha_2$ & $Ke^{\alpha_2 r}-1$
        \\
		\hline
		Coulomb & $\neq0$ & $0$ & $0$ & $(-\alpha_2 r-K)^{-1}$
		\\
		\hline
		harmonic & $0$ & $0$ & $>0$ & $\alpha _0 r+K$
		\\
        \hline
		quotient exponential & $\neq0$ & $\alpha_2$ & $0$ & $-e^{K+\alpha _2 r}/(e^{K+\alpha _2 r}-1)$
        \\
        \hline
	\end{tabular}
	\caption{
Some characteristic couplings belonging to the
family \eqref{solutions-exact-mapping}. Their Pekeris mappings correspond to
a Schr\"odinger equation with a non-minimal coupling $U(r)$, given by \eqref{mapping-exact}.}
\end{table}

\section{Conclusions}

We have presented the one-dimensional
and the three-dimensional relativistic equations for the Morse potential that result from
a generalized momentum operator provided with a deformed non-minimal coupling.
By means of the Pekeris approximation in the 3D case
we have converted the not exactly solvable
radial wave equation \eqref{structural-radial}
into the
Morse-like equation \eqref{Morse-radial-compact}, whose solutions
and energies are obtained by a mapping onto the one-dimensional
Morse problem
%Schr\"odinger equation
\eqref{Morse-SE},
corresponding to the vibrational states ($r\sim r_{\mathrm{e}}$).
We have recovered the non-relativistic energies of the $\textrm{S}$-wave states of the
$\textrm{H}_2$ in a very good agreement (Table 1) and we have shown that
the Pekeris approximation gives a good accuracy of the Coulomb and centrifugal terms (Fig. 1) within
the range $r_e/2 \leq r \leq 3r_e/2$.
%the effect of the total spin ($j=1/2+l$) in the Dirac case is to increase the dissociation energy
%with respect to the KG case (Fig. 1).
%

For the three-dimensional case and employing the Pekeris approximation \eqref{generelized-Pekeris},
we have seen that
the corrections of the spin and momentum angular contributions
to
the one-dimensional Morse energies \eqref{eigenergies-QMO} are contained in the effective angular frequency
$\Omega^2=\omega^2\left[+\delta\left(3+2(1+f(j,l))\frac{1+\alpha}{\alpha}+l(l+1)\frac{\alpha^2+3\alpha+2}{\alpha}\right)\right]$,
leading to the energy formula \eqref{eq:energy-recasted},
which is valid for the $\textrm{H}_2$, LiH, HCL and CO molecules when $\delta\ll1$ (Table 2).
%the $\textrm{H}_2$, LiH, HCL and CO molecules in the approximation of small $\delta\ll1$ (Table 2).
We have illustrated the spin effect
to the Dirac energies of the S-wave states
with the $\textrm{H}_2$ and LiH molecules, where an splitting in the energies is evidenced (Fig. 2).
Regarding the thermodynamical properties, for the molecules studied and
for a high energy electron in the non-relativistic and relativistic regimes
respectively, we have reported
Schottky effects in the heat capacity due to the finiteness of the allowed spectrum of the Morse potential (Figs. 3 and 4).
In the Dirac and KG systems, the Schottky peaks express
the screening of the spin contributions to the heat capacity (caused by thermal excitations), thus making them
appreciable at low temperatures in the non-relativistic regime or at high temperatures in relativistic particles (Table 3).

By revisiting the generalized Pekeris approximation \cite{Pekeris2},
we have established the functional form $f(x)=\gamma U^{-1}((x-1)/\gamma)-\gamma r_e$
of the expansion variable $y=f^{-1}(\gamma(r-r_e))$,
%of the Pekeris approximation
%in terms of an arbitrary non-minimal coupling $U(r)$,
and we have extended this perspective to the relativistic domain. Thus, given $U(r)$
we have a Pekeris mapping from the three-dimensional
KG and Dirac equations to a one-dimensional like
Schr\"odinger equation,
%, that we called \emph{Pekeris mapping},
given by Eqns.
\eqref{pre-general-pekeris} \eqref{structural-mapping} and \eqref{constants-mapping}.
We have illustrated the complexity of the Pekeris mapping
%according to the function $f(x)$,
for the 12-6-9 Lennard-Jones and the homographic-squared potentials.
Moreover, we have obtained the family of non-minimal couplings whose
Pekeris mapping becomes into a one-dimensional Schr\"odinger equation
provided with a minimal coupling \eqref{solutions-exact-mapping},
from which the tangent, Morse, Coulomb, harmonic and the quotient exponential
result to be special cases (Table 4).

\begin{comment}
From the best of our knowledge, this is the first time that the Morse potential is obtained by
the relativistic theory by its own, doing a simply canonical transformation of the position variable, for the 1D case,
and with a deformed linear potential, in the 3D case. In \cite{Costa_Filho-2013} the authors obtain the Morse potential
by first principles for the non-relativistic quantum mechanics, and with a similar idea,
we have obtained a good generalization in relativistic contexts with a difference
in the effective frequency $\Omega$,
as a result of the non-minimal coupling and the Pekeris approximation.
%

Our study reveals that generalized
momentum operators
can be extended
to relativistic contexts by means of canonical transformations and
by non-minimal couplings which represent a deformation of the
linear harmonic coupling, thus giving place to
the relativistic Morse-like oscillator by first principles.
Furthermore, this can applied to study energy levels,
S-wave states, relativistic dissociation energies, and for future works
the rotational states ($l\neq0$)
with the help of numerical simulations.
\end{comment}

%--------------------------------------\\
\section*{Acknowledgments}
	The authors acknowledge support received from the National Institute of Science
	and Technology for Complex Systems (INCT-SC),
	and from the CNPq and the CAPES (Brazilian agencies) at Universidade Federal da Bahia, Brazil.
%	\\

\section{Appendix
%: Generalized Pekeris approximation and radial equation
}

We have that
up terms of second order \cite{Pekeris-1933,Pekeris2}
\begin{eqnarray}\label{generelized-Pekeris}
&r_{\mathrm{e}}/r=
\left(1+\ln y /\gamma r_{\mathrm{e}}\right)^{-1}
\approx1-\frac{1}{\gamma r_{\mathrm{e}}}(y-1) \nonumber\\
&+\frac{2+\gamma r_{\mathrm{e}}}{2(\gamma r_{\mathrm{e}})^2}(y-1)^2
\nonumber\\
&(r_{\mathrm{e}}/r)^2=\left(1+\ln y /\gamma r_{\mathrm{e}}\right)^{-2}\approx 1-\frac{2}{\gamma r_{\mathrm{e}}}(y-1)\nonumber\\
&+\frac{3+\gamma r_{\mathrm{e}}}{(\gamma r_{\mathrm{e}})^2}(y-1)^2
\end{eqnarray}
with $y=e^{\gamma (r-r_{\mathrm{e}})}$ and the expansions are around $y=1$ ($r=r_{\mathrm{e}}$).
The effective potential of the radial equation \eqref{structural-radial} is
\begin{equation}\label{potential-radial}
U_{\mathrm{eff}}(r)=
\frac{m\omega^2}{2}U^2-\frac{\hbar\omega}{2}\frac{dU}{dr}
-[1+f(j,l)]\frac{\hbar\omega U}{r}+\frac{\hbar^2 l(l+1)}{2mr^2}.
\end{equation}
In order to adimensionalize variables we set $-\gamma r_e=\alpha>0$ and $\hbar/(m\omega r_e^2)=\delta>0$.
Thus, replacing $U=\frac{e^{\gamma(r-r_e)}-1}{\gamma}$ and the Pekeris approximated expressions of
$r_{\mathrm{e}}/r,(r_{\mathrm{e}}/r)^2$ of \eqref{generelized-Pekeris} in \eqref{potential-radial}
we can rewrite the effective potential $U_{\textrm{eff}}(r)$ as
\begin{equation}
U_{\textrm{eff}}(r)=\nonumber
\frac{m\Omega^2}{2\gamma^2}\left[e^{\gamma(r-r_{\textrm{eff}})}-1\right]^2+U_0
\end{equation}
where
\begin{eqnarray}\label{redefinition-constants}
&\Omega^2=\omega^2 A\left(1-\frac{B}{2A}\right)^2\nonumber
\\
&r_{\mathrm{eff}}=r_{\mathrm{e}}+\frac{1}{\gamma}\ln\left(1-\frac{B}{2A}\right)
\nonumber \\
&U_0=\frac{m\Omega^2}{2\gamma^2}\left(C-\frac{B^2}{4A}\right) \nonumber
\end{eqnarray}
and
\begin{eqnarray}\label{constants}
&A=1+\delta\left(2(1+f(j,l))+l(l+1)(3-\alpha)\right)\nonumber
\\
&B=-\delta\left[1-2\frac{1+f(j,l)}{\alpha}-2\frac{l(l+1)}{\alpha}\right]\nonumber\\
&C=\delta\left[(l+1)-1\right]\nonumber
\end{eqnarray}

%%%%%%%%%%%%%%%%%%%%%%%%%%%%%%%%%%%%%%%%%%%%%%%%%%%%%%%%%%%%%%%%%%%%%%%%%%%%%%%%%%%%%%%%%%%%%%%%%%%%%%%%%%%%
%%%%%%%%%%%%%%%%%%%%%%%%%%%%%%%REFERENCES%%%%%%%%%%%%%%%%%%%%%%%%%%%%%%%%%%%%%%%%%%%%%%%%%%%%%%%%%%%%%%%%%%%
%%%%%%%%%%%%%%%%%%%%%%%%%%%%%%%%%%%%%%%%%%%%%%%%%%%%%%%%%%%%%%%%%%%%%%%%%%%%%%%%%%%%%%%%%%%%%%%%%%%%%%%%%%%%

\end{document}